\documentclass[fleqn,usenatbib,onecolumn]{mnras}
\usepackage[utf8]{inputenc}
\usepackage[T1]{fontenc}
\usepackage{ae,aecompl}
\usepackage{graphicx}	
\usepackage{amsmath}	
\usepackage{amssymb}	
\usepackage{color}
\usepackage{ulem}
\usepackage{natbib}
\newcommand{\VEC}[1]{\vec {#1} }
\title[
Maximum gap in the magnetic field across crusts]
{How different is the magnetic field at the core-crust interface from that at the neutron star surface?:the range allowed in magnetoelastic equilibrium
}
 \author[Y. Kojima, S. Yoshida]{
  Yasufumi Kojima 
 \thanks{%
 E-mail: ykojima-phys@hiroshima-u.ac.jp}$^1$,
 Shijun Yoshida$^2$\\ 
 $^1${Department of Physics 
 Graduate School of Advanced Science and Engineering, Hiroshima University
}\\
{Higashi-Hiroshima 739-8526, Japan}\\
$^2${Astronomical Institute, Tohoku University, Sendai 980-8578, Japan}
}
 
\begin{document}
\maketitle 
%
\begin{abstract}
This study was focused on the investigation of a magnetic field penetrating from the core of a neutron star to its surface.
The range of possible field configurations
in the intermediate solid crust is less limited owing to the elastic force acting on the force balance.
When the Lorentz force is excessively strong, the magnetoelastic equilibrium does not hold, and thus, the magnetic field becomes constrained.
By numerically solving for the magnetoelastic equilibrium in a thin crust,
the range of the magnetic field at the core--crust interface was determined,
  while assuming the exterior to be fixed as a dipole in vacuum.
The results revealed that
the toroidal component should be smaller than the poloidal component at the core--crust interface
for the surface dipole, $B_{0} > 2.1 \times  10^{14}$G.
Consequently, a strong toroidal field, for example, $B \sim 10^{16}$G, as suggested by free 
precession of magnetars
should be confined to a deep interior core and
should be reduced to $B \sim 10^{14}$G at the bottom of the crust. 
The findings of this study provide insights into the interior field structure of magnetars.
Further investigations on more complicated geometries with higher multipoles and exterior magnetosphere are necessary.
\end{abstract}
%
%
\section{Introduction}
A neutron star is surrounded by a thin, light shell called the crust.
Its thickness is less than 0.1 times the stellar radius,
while its mass is less than 0.01 of the total mass~\citep[for example][]{1983bhwd.book.....S}.
Therefore, the outer layer can be neglected in the first approximation when the entire stellar model is considered. 

The ions in the crust constitute a Coulomb plasma, which is 
 typically formed in crystals ~\citep[for example][]{2008LRR....11...10C}.
In contrast to fluids which are the primary material form in all stars, elastic media 
may be important for the crust-quake that occur in magnetars, which are strongly magnetized neutron stars~\citep[][for a seminal paper]{1992ApJ...392L...9D,1995MNRAS.275..255T}.
Sudden crust breaking can produce a magnetar outburst and/or a fast radioburst~\citep{2015MNRAS.449.2047L,2016ApJ...833..189L,2018MNRAS.480.5511B,2019MNRAS.488.5887S}.
The accumulation of shear stress toward crustal fractures owing to the
the evolution of the Hall magnetic field is studied~\citep{2022ApJ...938...91K,2023ApJ...946...75K}.
Recent theoretical works demonstrate that
the elastic force plays a crucial role on the dynamical stability
~\citep{2020MNRAS.499.2636B}, and on 
static equilibrium models interiorly possessing strong magnetic field 
~\citep[][]{2021MNRAS.506.3936K,2022MNRAS.511..480K,
2022MNRAS.516.5196F}.

The following question arises for a magnetic field penetrating from the inner core to the exterior vacuum:
How much difference in magnitude is allowed between the magnetic field
at the core--crust interface and that at the surface?
This depends on the electric current distribution in the crust.
However, determining the possible range is useful for further studies.
Moreover, boundary conditions are imposed at the core--crust interface without detailed models in the crust
when considering magnetic field in the core.
%

 Free precessions of spinning magnetars observed in 4U 0142+63
~\citep{2014PhRvL.112q1102M,2019PASJ...71...15M},
in 1E 1547-54
~\citep{2016PASJ...68S..12M,2021MNRAS.502.2266M},
and in SGR 1900+14
~\citep{2021ApJ...923...63M}
suggest a strong toroidal component relevant to deformation.
The toroidal component inside these sources is $\sim 10^{16}$G,
whereas the surface dipole field is $(1.3 - 7)\times 10^{14}$G.
Such strong components may affect the magnetic structure of the crust unless
it is confined to a deep interior.
Our study provides useful information regarding magnetic field
at the inner boundaries of the crust.

The permissible range depends on the configuration and strength of the magnetic field.
The field configuration is arbitrary in the very weak case.
As the field strength for the fixed configuration increases, 
the Lorentz force significantly affects force balance.
We consider equilibrium with the Lorentz and elastic forces, because the latter in the crust is important for extending the range of the magnetic field configuration.
However, the magnetoelastic force balance is disrupted 
when the elastic deformation exceeds a certain threshold.
Thus, the constraint leads to an upper limit for the magnetic field strength.

We require the following minimum conditions for
magnetic field geometry of the crust. 
The electric current in the thin layer is assumed to be described by a smooth function that matches the conditions on both sides of the spherical shell.
This case is simple and natural for a thin crust.
For a more complicated distribution, the magnetic field is tangled such that the contribution of the elastic force increased accordingly.
Therefore, the constraints become more severe.
%

The remainder of this paper is organized as follows. The models and equations are discussed in Section 2. 
In Section 3, we calculate the magnetoelastic equilibrium for
the minimum required electric currents and
derive a change in the magnetic field across the crust.
Finally, Section 4 concludes the study.

\section{Mathematical formulation}
\subsection{Magnetic field}
We consider the magnetic field 
from the core--crust interface at $r_{c}$ to 
surface at $R\equiv r_{c}+\Delta r$.
The magnetic field was limited to an axially symmetric dipole, which is
expressed in spherical coordinates $(r, \theta, \phi)$ as follows:
\begin{equation}
  B_{r}=  \frac{2g}{r^2}\cos\theta,~~
  B_{\theta} =-\frac{g^{\prime}}{r}\sin\theta,~~
  B_{\phi}=\frac{s}{r}\sin\theta,
 \label{dipolB.eqn}
\end{equation}
where $g$ and $s$ are functions of $r$ and
 prime $^\prime$ denotes the derivative of $r$.
In an external vacuum $(r\ge R)$, these functions are expressed as
\begin{equation}
g=\frac{B_{0}R^3}{2r}, ~~s=0,
\label {extBfld.eqn} 
\end{equation}
where $B_{0}$ is the field strength of the magnetic pole
on the surface $r=R$.
%

\begin{figure}\begin{center}
\includegraphics[scale=0.5]{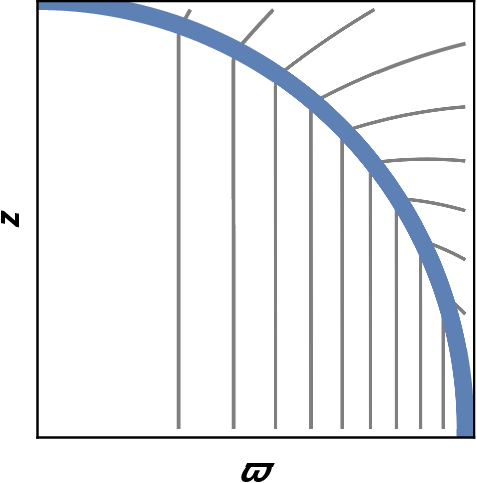}
\includegraphics[scale=0.5]{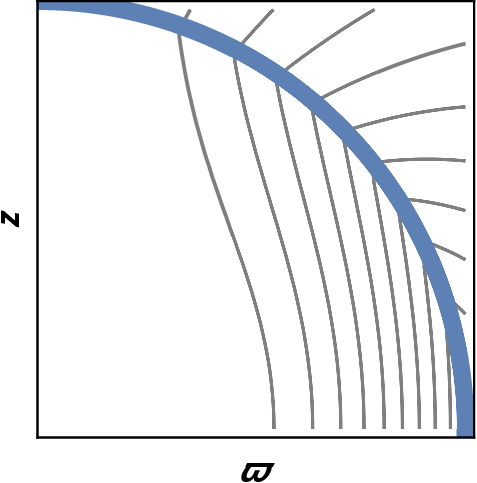}
\includegraphics[scale=0.5]{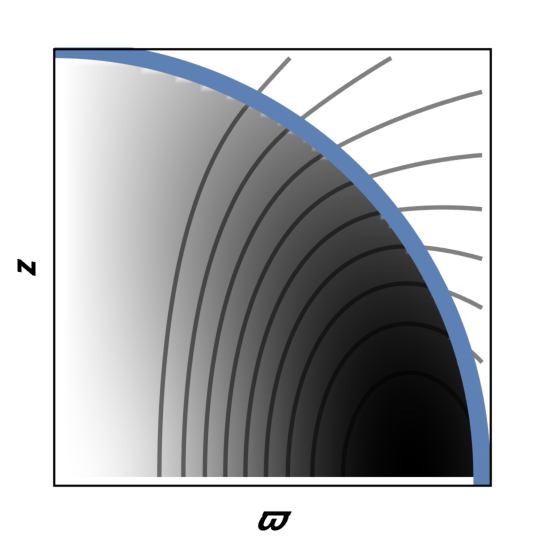}
\caption{ 
  \label{Fig0}
Three examples of magnetic fields penetrating the crust, denoted by the thick blue
arc.
Polodial field lines and greyscale contour map of toroidal field
$B_{\phi}$ are shown in meridian plane 
($\varpi =r \sin\theta$, $z=r\cos\theta$).
The left and middle panels exhibit $\Delta B_{\theta}\ne 0$
across the crust, while the right panel shows $\Delta B_{\phi}\ne 0$.
}
\end{center}\end{figure}

When the thickness $\Delta r$ is small
$\Delta r \ll r_{c}$, the change in 
of the magnetic field along a fixed polar angle $\theta$ can be approximated as follows: 
\begin{align}
 &\Delta B_{r}\equiv B_{r}(R,\theta) - B_{r}(r_c,\theta) 
\equiv  B_{0}\Delta \alpha \cos\theta,
 \label{DiffBr.eqn}
 \\
 &\Delta B_{\theta}\equiv B_{\theta}(R,\theta) - B_{\theta}(r_c,\theta)
 \equiv B_{0} \epsilon_{g} \sin\theta,
 \label{DiffBt.eqn}
 \\
&\Delta B_{\phi}\equiv B_{\phi}(R,\theta) - B_{\phi}(r_c,\theta) 
\equiv B_{0} \epsilon_{s}\sin\theta,
 \label{DiffBx.eqn}
\end{align}
where $\Delta \alpha$, $\epsilon_{g}$ and $\epsilon_{g}$
depends on the current distribution in the shell region between $r_{c}$ and $R$,
When we regard the crust as an infinitesimally thin
layer $\Delta r \to 0$,
$\Delta B_{r}$ vanishes owing to the continuity
derived by ${\vec{\nabla}}\cdot {\vec{B}}=0$.
However, the discontinuity in tangential components is allowed, 
 and $\Delta B_{\theta}$ and $\Delta B_{\phi}$ may be finite corresponding to the surface current.
 Three examples of the magnetic field with a discontinuity in the tangential component
are shown in Fig.~\ref{Fig0}.
The exterior field is the same for all of them, but
$B_{\theta}(r_c,\theta)$ and $B_{\phi}(r_c,\theta)$ 
at the bottom of the crust are different.
 The current density in the crust
 may be extremely strong in several cases
 in the thin limit approximation, and hence the
 resulting Lorentz force may violate the magnetoelastic equilibrium.

%
We consider a model for
the tangential components of the magnetic field in the crust.
The radial functions $g^{\prime}$ and $s$ in eq.~(\ref{dipolB.eqn})
can be approximated using the linear function of
$R-r$ within the range  $r_c \le r \le R$.
The radial function $g$ is a quadratic equation of $R-r$.
By using the Amp{\'e}re--Bio--Savart's equation
$\VEC{\nabla}\times {\VEC{B}}=4\pi {\VEC{j}}/c$,
that is matched with eqs.~(\ref{extBfld.eqn})--(\ref{DiffBx.eqn}),
the radial functions $g$ and $s$ are determined as follows:
\begin{equation}
g = \frac{1}{2} B_{0}
\left[ R^2+R(R-r) 
-\left(  \frac{1}{2}+\frac{r_{c}\epsilon_{g}}{\Delta r}
\right)(R-r)^2
\right],
\label{gfunction2.eqn}
\end{equation}
\begin{equation}
s=-\frac{B_{0}r_{c}\epsilon_{s}}{\Delta r}(R-r).
\label{sfunction1.eqn}
\end{equation}
The coefficient $\Delta \alpha$ is obtained as:
\begin{equation}
\Delta \alpha =\frac{\Delta r}{r_{c}}
\left[\epsilon_{g} -3 -\frac{3\Delta r}{2r_{c}}
\right].
\label{deltaalpha.eqn}
\end{equation}
The coefficients $\epsilon_{g}$ and $\epsilon_{s}$
in eqs.~(\ref{DiffBt.eqn})--(\ref{DiffBx.eqn}) are
related to the integration of the electric current in the spherical shell as follows:
\begin{align}
 & \frac{4\pi }{c}\int_{r_c} ^{R} r j_{\phi} dr=
  B_{0} r_{c}\left(\epsilon_{g} +  \frac{\Delta r}{2r_{c}}
  \right) \sin\theta,
 \\
 & \frac{4\pi }{c}\int_{r_c} ^{R}  rj_{\theta}dr=
 -B_{0} r_{c}\epsilon_{s}\sin\theta.
\end{align}

For the magnetic field penetrating from inner core to the
exterior vacuum, estimating the allowed difference in each magnetic field component is crucial.
As shown by eq.~(\ref{deltaalpha.eqn}),
  $\Delta B_r/B_{0}$ in the radial component,
is of the order $\sim \Delta r /r_{c}$ and is consistent with the continuity 
within the limit $\Delta r \to 0$.
However, $\Delta B_\theta$ and $\Delta B_\phi$ 
are not simple, and some calculations are necessary.
These values were determined by considering the force balance in the crust.
Thus $\epsilon_{g}$ and $\epsilon_{s}$ 
in eqs.~(\ref{DiffBt.eqn})--(\ref{DiffBx.eqn}),
which are related with 
the surface current in the thin limit,
are constrained by the elastic limit of the equilibrium.
%

\subsection{Magnetoelastic equilibrium in crust}
The Lorentz force ${\VEC{f}}$ for the dipolar field (\ref{dipolB.eqn}) is expressed as
\begin{equation}
4\pi{\VEC{f}}
 = \frac{4\pi j_{\phi} }{cr\sin \theta} {\VEC{\nabla}} 
 (g \sin^2 \theta) 
 - \frac{s}{r^2} {\VEC{\nabla}}(s \sin^2 \theta)
 +\frac{2}{r^2} (gs^{\prime}- g^{\prime}s)\sin \theta \cos \theta
 ~{\VEC{e}}_{\phi},
\label{LFpolcomp.eqn}
\end{equation}
where
${\VEC{e}}_{\phi}$ is a unit vector in the $\phi$-direction.
The acceleration vector due to the Lorentz force is generally decomposed into
irrotational and solenoidal components.
The irrotational component, which is expressed by the gradient of a scalar function, may be balanced by a change in pressure and gravity.
The ratio of the Lorentz force to these dominant forces is typically ($10^{-4}$--$10^{-7}$)$\times(B_{0}/10^{14}{\rm G})^2$ depending on the stellar radius. Therefore, the Lorentz force may be considered as a small perturbation of a spherical stellar structure. 
However, there is no solenoidal component of gravity and pressure in barotropic stars.
This fact strongly constrains the barotropic equilibrium models.
Static or stationary axially symmetric models have been calculated under various conditions using various methods
~\citep[for example][]{2005MNRAS.359.1117T,2006ApJS..164..156Y,
2006ApJ...651..462Y,2009MNRAS.395.2162L,
2010A&A...517A..58D,
2013MNRAS.432.1245F,2013MNRAS.434.2480G,
2015ApJ...802..121A}.
The results revealed that 
the toroidal magnetic field was much smaller than that of the poloidal magnetic field.
A similar situation is prevalent for realistic relativistic models~\citep[for example][]{2009MNRAS.397..913C,
2019PhRvD.100l3019U,2023PhRvD.107j3016U}.
The stratified structure of density and pressure, which results in solenoidal acceleration, is  a critical factor for stable
mixed poloidal and toroidal magnetic fields~\citep[for example][]{2009A&A...499..557R,2012MNRAS.420.1263G,2012MNRAS.424..482L,2013MNRAS.433.2445A,2019PhRvD..99h4034Y},
 as reported in an earlier study~\citep{1980MNRAS.191..151T}.
In this study, we did not consider 
stratification; however, the elasticity in the neutron star crust was incorporated. 
A ”curl” of the Lorentz force is balanced with the elastic force ${\VEC{h}}$.
The force balance with the extra force
allows different structures,
although the elastic force is significantly weaker than
 the dominant forces and its typical maximum magnitude ratio is $10^{-4}$.

A set of approximated equations
relevant to magnetoelastic equilibrium is
expressed as
\citep{2021MNRAS.506.3936K,2022MNRAS.511..480K}
\begin{align}
& ({\Vec f}+{\Vec h})_{\phi}=0, 
 \label{eq:elastic_phi}
 \\
& [{\nabla}\times \rho^{-1}({\Vec f}+{\Vec h})]_{\phi}=0,
  \label{eq:rot_elastic}
\end{align}
where $\rho$ denotes the mass density, which is a function of  $r$.
We considered only the azimuthal component in eq.~(\ref{eq:rot_elastic}) because the other poloidal components vanish according to eq.~(\ref{eq:elastic_phi}) and axial symmetry ($\partial_{\phi}=0$).

The {\it i}th component  $h_{i}$ is expressed by the shear modulus $\mu$ and
elastic displacement $\xi_{i}$ as follows:
\begin{align}
& h_{i}= {\nabla}_{j}(2 \mu \sigma _{i} ^{j}),
\\
& \sigma _{ij}= 
\frac{1}{2}({\nabla}_{i} \xi _{j} +{\nabla}_{j} \xi _{i}),
\end{align}
where the incompressible motion 
${\VEC{\nabla}} \cdot {\VEC{\xi}}=0$
is assumed in the expression of shear stress tensor $\sigma _{ij}$.
The elastic displacement induced by
the Lorentz force ${\vec{f}}=c^{-1}{\vec{j}}\times{\vec{B}}$ in eqs.~(\ref{eq:elastic_phi})--(\ref{eq:rot_elastic}) is expressed by
the Legendre polynomials $P_l(\cos\theta)$ with $l=2$ only,
because the angular dependence of
both ${\vec{j}}$ and ${\vec{B}}$ is given by 
$l=1$(see eq.~(\ref{LFpolcomp.eqn})).
We explicitly write the displacement, which
satisfies the incompressible conditions, as follows: 
\begin{equation}
\xi_{r}=\frac{6x_{2}}{r^2}P_{2},
~~
\xi_{\theta}=\frac{x_{2}^{\prime}}{r}P_{2, \theta},
~~
\xi_{\phi}=-rk_{2}P_{2, \theta}(\theta),
\end{equation}
where $x_2(r)$ and  $k_2(r)$ are the radial functions.

The crust is limited to the inner crust, where the mass density ranges from $\rho_c = 1.4 \times 10^{14} \mathrm{g~cm^{-3}}$ at the core--crust boundary $r_c$ to the neutron drip density, 
$\rho_1 = 4 \times 10^{11}\mathrm{g~cm^{-3}}$ at $R$. 
We ignored the outer crust with extremely small thickness and considered the exterior region as a vacuum. 
We assumed that the shear modulus is approximately proportional to the mass density such that it depends on the radial coordinate $\mu=\mu(r)$ as follows: 
\begin{equation}
\frac{\mu}{\mu_{c}}= \frac{\rho(r)}{\rho_c}
=\left[1-\left(1-\left(\frac{\rho_1}{\rho_{c}}\right)^{1/2}
 \right)\left(\frac{r-r_{c}}{\Delta r} \right)\right]^2 .
   \label{eq:mu}
\end{equation}
where $\mu_c =10^{30}~{\rm{erg}}~{\rm{cm}}^{-3}$ is the shear modulus at 
the core--crust interface~\citep[for a reasonable approximation]{2008LRR....11...10C,2019MNRAS.486.4130L}.

Using eqs.~(\ref{gfunction2.eqn})--(\ref{sfunction1.eqn})
eq.~(\ref{eq:elastic_phi}) is reduced to the following second-order differential equation:
\begin{equation}
 (\mu r^{4} k_{2}^{\prime})^{\prime} -4\mu r^{2} k_{2}
+a_2=0,
   \label{klexpd.eqn}   
\end{equation}
where 
\begin{equation}
a_2=
\frac{B_0 ^2}{24\pi}
\frac{ \epsilon_{s} r_c}{\Delta r}
\left[
2R^2+(R-r)^2+2(R-r)^2
\frac{\epsilon_{g} r_c}{\Delta r}
\right].
   \label{kla2.eqn}   
\end{equation}
Equation (\ref{eq:rot_elastic}) can be reduced to a fourth-order differential equation as follows:
\begin{equation}
  \left[\frac{(\mu y_{2})^{\prime}}{\rho}
+2\left(\frac{\mu^{\prime}}{\rho r}\right) x_{2}^{\prime}
\right]^{\prime}
-\frac{6}{r^2}
\left[ \frac{\mu y_{2}}{\rho}
+2\left(\frac{\mu^{\prime}}{\rho}\right)^{\prime}x_{2}
\right] +b_2=0,
\label{glexpd.eqn}
\end{equation}
\begin{equation}
    x_{2}^{\prime\prime} -\frac{6}{r^2}x_{2}+y_{2}=0,
\label{flexpd.eqn}
\end{equation}
where
\begin{align}
b_2&=\frac{B_0 ^2}{24\pi}
\bigg[
\left\{\frac{3R^2-r^2}{r_c}
-\frac{2(R-r)^2}{r_c}
\left( \frac{ \epsilon_{g} r_c}{\Delta r} \right)
\right\}\left(\frac{1}{\rho r} \right)^\prime
\left( \frac{ \epsilon_{g} r_c}{\Delta r} \right)
\nonumber
\\
&~~~+
4\left\{
\frac{R-r}{\rho r^2}
-(R-r)^2\left(\frac{1}{\rho r^2} \right)^\prime
\right\}
\left( \frac{ \epsilon_{s} r_c}{\Delta r} \right)^2
\bigg].
\label{glb2.eqn}
\end{align}

The source terms $a_2$ and $b_2$ in eqs.~(\ref{kla2.eqn}) and (\ref{glb2.eqn}) are derived from the Lorentz force.
Because of the thin shell $(r\approx R)$,
we neglected the terms proportional to $R-r$ in $a_2$ and $b_2$.
Consequently, the dependence of elastic displacement was clear.
In case of the axial displacement, 
$|\xi_{\phi}| \propto a_{2} \propto  \epsilon_{s}$, whereas for polar displacement, 
$|\xi_{p}| \propto b_{2} \propto \epsilon_{g}$,
Because the star is assumed to be spherically symmetric, the boundary conditions for eqs. (\ref{klexpd.eqn}), (\ref{glexpd.eqn}), and (\ref{flexpd.eqn}) are given by the force balance across the surfaces at $r_c$ and $R$. Thus, the shear stress tensors  $\sigma_{ri} ~(i=r,\theta, \phi)$ vanish because the other stresses of the fluid and magnetic field are assumed to be continuous.
The boundary conditions for the radial functions $k_{2}$, 
$x_{2}$ and $y_{2}$ at $r_{c}$ and $R$ are explicitly written as follows:
\begin{align} 
&  k_{2}^{\prime} =0,
  \label{bcT13}
  \\
  & \left(r^{-2}x_{2} \right)^{\prime}
 = 0,
   \label{bcT11}
 \\
&
2r x_{2} ^{\prime} -12x_{2}  +r^{2} y_{2} =0.
\label{bcT12}
\end{align}

\section{Numerical Results}
\subsection{Spatial shear--distribution}

\begin{figure}\begin{center}
      \includegraphics[width=0.8\columnwidth]{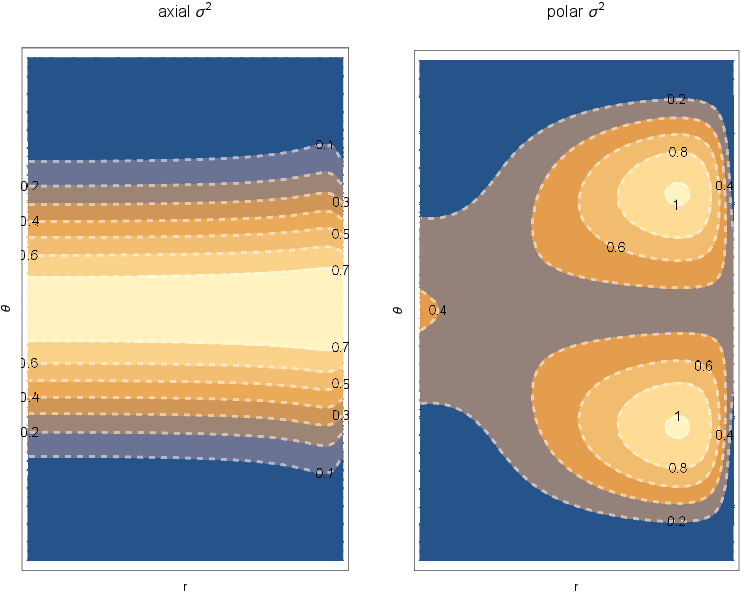}
\caption{ 
  \label{Fig1}
Contour map in $r-\theta$ plane of the magnitude $\sigma^2$ normalized by the maximum.
The crust region is given by
$r_{c} \le r \le R$ and $0\le \theta \le \pi $.
The left panel is $\sigma_{\rm{ax}}^2$ for
axial displacement, while
the right is $\sigma_{\rm{pl}}^2$ for polar displacement.
}
\end{center}\end{figure}


%
The magnitude of the shear stress is expressed as
\begin{equation}
\frac{1}{2}\sigma_{ij}\sigma^{ij}
=(\sigma_{\rm{ax}})^{2}+ (\sigma_{\rm{po}})^{2} ,
   \label{criterion}
\end{equation}
where the total is split
to the axial paper $\sigma_{\rm{ax}}$ and poloidal part
 $\sigma_{\rm{po}}$, because
 they were determined independently. 
The former is caused by $\xi_\phi \propto \epsilon_{s}$
and $(\xi_r, \xi_\theta)  \propto \epsilon_{g}$.
Figure~\ref{Fig1} illustrates the spatial
distribution of  $\sigma^2$ in
a range of
$r_c \le r\le R$ and $0\le \theta \le \pi$.
The profile does not change with the thickness $\Delta r$.
\cite{2022MNRAS.516.5196F} calculated  
$\sigma_{\rm{ax}}$ and $\sigma_{\rm{po}}$
to obtain a more elaborate current distribution model 
in the crust.
The profiles of $\sigma_{\rm{ax}}$ were almost the same.
However, the value of $\sigma_{\rm{po}}$ slightly differed from that of
\cite{2022MNRAS.516.5196F},
which contained two peaks in the radial direction.
This complicated structure originates from the electric current distribution
adopted in the models.
%

\subsection{Elastic equilibrium range}
The elastic equilibrium is possible
when the magnitude of the entire crust is less than
a threshold value $\sigma_{c}$ (the Mises criterion
~\citep{1969imcm.book.....M,2000MNRAS.319..902U}),
where $\sigma_{c}$ is the number  
$\sigma_{c} \approx 10^{-2}-10^{-1}$
~\citep[e.g.][]{2009PhRvL.102s1102H,2018PhRvL.121m2701C,2018MNRAS.480.5511B}. 
As both $\sigma_{\rm{ax}}$ and $\sigma_{\rm{po}}$
depend on the position, the equilibrium conditions are  
  $\sigma_{\rm{ax}} ^{\rm{max}} \le \sigma_{c}$ and
  $\sigma_{\rm{po}} ^{\rm{max}} \le \sigma_{c}$.
These conditions can be written as follows because  
$ \sigma_{\rm{ax}} \propto |\xi_{\phi}| \propto  |\epsilon_{s}|$ and 
$ \sigma_{\rm{po}} \propto |\xi_{p}| \propto  |\epsilon_{g}|$.
\begin{equation}
|\epsilon_{g}| 
\le \epsilon_{g} ^{\rm{max}}
\equiv  N_{*} F_{g},
~~
|\epsilon_{s}| 
\le \epsilon_{s} ^{\rm{max}}
\equiv N_{*} F_{s},
\label{upperlimit.eqn}  
\end{equation}
where $F_g$ and $F_s$ are
 numerically obtained for a model with thickness $\Delta r$, and
$N_{*}$ is a dimensionless normalization constant expressed as
 \begin{equation}
N{*} \equiv
\frac{4\pi \mu_{c} \sigma_{c}}{B_{0} ^2}
\approx   1.3 \times 10^{2}
\times \left(\frac{ \sigma_{c}}{0.1}\right)
 \left(\frac{\mu_{c} }{10^{30}{\rm erg~cm}^{-3}}  \right)
 \left(\frac{B_{0} }{10^{14}{\rm G}}  \right)^{-2} .
 \label{normalizationN*.eqn}
\end{equation}

\begin{figure}\begin{center}
      \includegraphics[width=0.85\columnwidth]{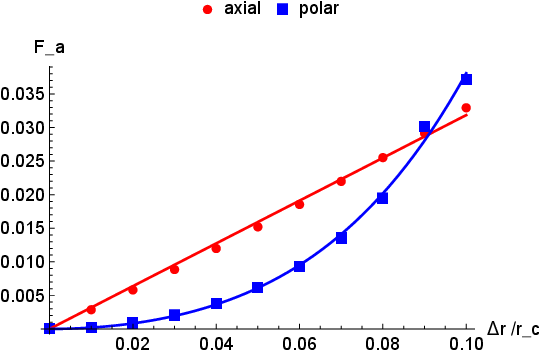}%
\caption{ 
 \label{Fig2}
The maximum deviation 
$F_{a}=\epsilon_{a} ^{\rm{max}} N_{*}^{-1}$
$(a=s,g)$ are plotted as
a function of thickness $\Delta r /r_{c}$.
The axial case was approximated by
a linear function of $x =\Delta r/(0.1 r_{c})$, 
whereas the polar was approximated by
$F_{g}=\beta_2 x^2 + \beta_4 x^4$.
}
\end{center}\end{figure}


%
The numerical results for $F_g$ and $F_s$ are plotted as
a function of $\Delta r /r_{c}$, as shown in Fig.~\ref{Fig2}.
The function $F_s$ was almost proportional to the
thickness, whereas $F_g$ changed more significantly with $\Delta r/r_{c}$.
We searched for an empirical fitting curve for these functions.
Assuming polynomials, the coefficients were obtained as the best-fit values.
The function for the axial part was approximated well by $F_{s}=\alpha_1 x$,
where $x =\Delta r /(0.1 r_{c})$ and $\alpha_1 =3.2\times 10^{-2}$.
For the polar part, it was $F_{g}=\beta_2 x^2 + \beta_4 x^4$, where 
$\beta_2 \approx 2.0\times 10^{-2}$, and $\beta_4 \approx 1.8\times 10^{-2}$.
%

Our model shows that the upper limits 
$\epsilon_{g} ^{\rm{max}}(\propto F_{g})$ and 
$\epsilon_{s} ^{\rm{max}}(\propto F_{s})$
approach zero as $ \Delta r \to 0$, although
 both the electric current and 
Lorentz force $f$ diverge as $(j,f) \propto (\Delta r)^{-1}$.
This convergence to zero results from the severe constraint on the elastic equilibrium.
Shear stress $\sigma_{\rm{ax}}$
is proportional to $\sigma_{\rm{ax}} \propto \xi_{\phi} \propto \epsilon_{s}/\Delta r$ and
 is limited to a finite value $ \sigma_{c}$.
Therefore, the maximum value of $\epsilon_{s}$ linearly depends on $\Delta r$, while 
the polar component $\sigma_{\rm{po}}$ is slightly different. 
An additional term 
$[1/(\rho r)]^{\prime} \sim (\rho_{1} R \Delta r)^{-1}$
is involved in the source term; therefore, 
 $\sigma_{\rm{po}} \propto \xi_{p} \propto \epsilon_{g}/(\Delta r)^2$.
The dependence on $\Delta r$
became quadric, that is,
 $\epsilon_{g}\propto (\Delta r)^2$
 in the small region $\Delta r$.
%

The change in the tangential component of the magnetic field
is formally expressed as eq.~(\ref{upperlimit.eqn}). However, it is actually less constrained, 
unless the magnetic field is sufficiently strong, that is, $\sim 10^{14}$G.
For example, for typical pulsars with $B_{0}=10^{12}$G,
the condition is
$|\Delta B_{\theta}/B_{0}| < 10^{5}$ and
$|\Delta B_{\phi}/B_{0}| < 10^{5}$;
there is no substantial limit to the tangential components.
For example, 
the direction and magnitude of $B_{\theta}$
at the core-crust interface may completely differ from those at the surface.
However, the magnetoelastic equilibrium range is
limited to the strong magnetic-field regime. 
This constraint is meaningful, 
because the normalization factor (\ref{normalizationN*.eqn})
decreases as $B_{0}$ increases.
$|\Delta B_{\theta}/B_{0}|<1$ and $|\Delta B_{\phi}/B_{0}|<1$
for magnetars with $B_{0}>2.1\times 10^{14}$G  assuming 
crustal thickness, $\Delta r/r_{c}=0.1$.
That is, the tangential components 
at the core-crust interface are 
within small extrapolation range from those at the surface.
Coefficient $|\epsilon_{s}|$ represents the ratio
 of the toroidal to poloidal components
 at the core--crust interface; 
  $|\epsilon_{s}|=|\Delta B_{\phi}/(B_{0}\sin \theta)|$
  $\approx |B_{\phi}(r_c,\pi/2)/B_{r}(r_c,0)|$.
The toroidal component should be smaller than 
poloidal components for $B_{0}>2.1\times 10^{14}$G.
The ratio decreases further with an increase in $B_{0}$.
Therefore, a strong toroidal component 
$B_{\phi} > 10^{15}$G 
should be confined to the interior of magnetars $r<r_c$.

 \section{Summary and Discussion}
We studied the change in the dipolar
magnetic field between
inner and outer boundaries in neutron star crusts.
An arbitrary configuration is allowed under an extremely weak magnetic field.
The Lorentz force significantly affects the
force balance as the field strength increases.
However, even for magnetars with the strongest field strength,
the maximum magnitude of the Lorentz force is $10^{-4}$
times smaller than those of pressure and gravity. 
%

The acceleration resulting from
these dominant forces are irrotational in the barotropic case, whereas the Lorentz force generally leads to
both irrotational and solenoidal components.
Therefore, at the equilibrium, the magnetic field is highly constrained,
or a different force exerts a force balance. 
In this study, we considered that the elastic force
was balanced by the solenoidal part of the Lorentz force. 
The elasticity of the solid crust increases the allowable range of the magnetic field.
%

The electric current in the thin layer is expressed as
using a simple smoothing function determined by
 the conditions at two surfaces of spherical shells. 
However, the magnetoelastic force balance is disrupted 
when the elastic deformation exceeds a certain threshold.
The elastic limit based on the Mises criterion 
constrains the magnetic-field strength.
%

Our model demonstrates that the difference
between magnetic fields
in the core crust and on the surface
vanishes with decreasing thickness,
although the electric current density 
diverges to the zero limit.
The discontinuities of the tangential components are generally allowed as a mathematical boundary condition of the magnetic field across an infinitesimally thin layer; however, the magnetoelastic force balance in the shell with a finite thickness always constrains the change.
The allowable range of the magnetic field decreases, although the Lorentz force increases with the decreasing thickness.
%

For the thickness values of $\Delta r /R \approx 0.02-0.1$, which is a realistic range for
the crust thicknesses of neutron stars,
the change in the magnetic field was highly constrained in 
the strong regime $B\sim 10^{14}$G, where
it was difficult for the elastic force to control the Lorentz force. 
Consequently, the magnetic field at the core--crust interface 
did not differ significantly from those of the
surface owing to its small thickness. 
The field strength
$B_{\phi}$  of the toroidal component at the core--crust interface was less than that of the surface dipole field $B_{0}$ 
when $B_{0}> 2.1 \times 10^{14}$G,
because the toroidal field is zero in an external vacuum.
The strong toroidal component $\sim 10^{16}$G
in the magnetar interior 
can be inferred from the observed free precession~\citep{2014PhRvL.112q1102M,2016PASJ...68S..12M,
2019PASJ...71...15M,2021MNRAS.502.2266M,2021ApJ...923...63M}; however,
the strength should be reduced to $\sim 10^{14}$G
at the bottom of crust.
Conversely, a strong component should be confined to the core.
Otherwise, an elaborate model would be required.
In particular, a superconducting core is crucial.
For example, 
the interior magnetic--field configuration
significantly depends on a ratio of 
the field strength at the crust–core boundary
to a characteristic value $H_{\rm{cl}}\sim 10^{15}$G
for the type-II superconductor
~\citep{2013PhRvL.110g1101L,2014MNRAS.437..424L}.

Studies focused on the development of magnetar models are ongoing, and further investigations are required
to examine the magnetic deformation of magnetars
~\citep[for example][and references therein]{2021MNRAS.503.2764F,2021A&A...654A.162S},
which is relevant to
the detection of 
 continuous gravitational waves,
 and account for the electromagnetic phenomena associated with outbursts
~\citep[for example][]{2023MNRAS.523.4089S}.
The crust has not appropriately been considered in most models.
Although the geometry and strength of the internal magnetic fields remain largely unknown, the present results 
are a step towards gaining insights into the 
condition at the core--crust interface from the stellar surface.
%

\section*{Acknowledgments}
 This work was supported by JSPS KAKENHI Grant Numbers JP19K03850, JP23K03389(YK).

 \section*{DATA AVAILABILITY}
%
Numerical code and data underlying this article will be shared on reasonable request 
to the corresponding author.
%

 \bibliographystyle{mnras}
 \bibliography{kojima23Aug} 

\end{document}